%% file: paper.tex
\documentclass[conference]{IEEEtran}
\IEEEoverridecommandlockouts

\usepackage{notoccite}
\usepackage{graphics}
\usepackage{multirow}
\usepackage[colorinlistoftodos]{todonotes}
\usepackage{caption}
\usepackage{lipsum}
\usepackage{mwe}
\usepackage{booktabs} 
\usepackage{todonotes}
\usepackage{soul}
\usepackage[breaklinks=true]{hyperref} 
\usepackage[euler]{textgreek} 
\def\BibTeX{{\rm B\kern-.05em{\sc i\kern-.025em b}\kern-.08em
    T\kern-.1667em\lower.7ex\hbox{E}\kern-.125emX}}

\begin{document}

\title{An Energy-Efficient Artefact Detection Accelerator on FPGAs for Hyper-Spectral Satellite Imagery}

\author{
 \IEEEauthorblockN{Cornell Castelino\IEEEauthorrefmark{1}, Shashwat Khandelwal\IEEEauthorrefmark{1},
 Shanker Shreejith\IEEEauthorrefmark{1},
 Sharatchandra Varma Bogaraju\IEEEauthorrefmark{2}}
 \IEEEauthorblockA{\IEEEauthorrefmark{1}Reconfigurable Computing Systems Lab, Electronic \& Electrical Engineering
 Trinity College Dublin, Ireland.}
 \IEEEauthorblockA{\IEEEauthorrefmark{2}Faculty of Computing, Ulster University, Jordanstown Campus, Newtownabbey, U.K}
 \{castelic, khandels, shankers\}@tcd.ie, s.bogaraju@ulster.ac.uk
}

\maketitle
\begin{abstract}

Hyper-Spectral Imaging (HSI) is a crucial technique used to analyse remote sensing data acquired from Earth observation satellites. 
The rich spatial and spectral information obtained through HSI allows for better characterisation and exploration of the Earth's surface over traditional techniques like RGB and Multi-Spectral imaging on the downlinked image data at ground stations.
In some cases, these images do not contain meaningful information due to the presence of clouds or other artefacts, limiting their usefulness.
Transmission of such artefact HSI images leads to wasteful use of already scarce energy and time costs required for communication. 
While detecting such artefacts prior to transmitting the HSI image is desirable, the computational complexity of these algorithms and the limited power budget on satellites (especially CubeSats) are key constraints. 
This paper presents an unsupervised learning-based convolutional autoencoder (CAE) model for artefact identification of acquired HSI images at the satellite and a deployment architecture on AMD's Zynq Ultrascale FPGAs.
The model is trained and tested on widely used HSI image datasets: Indian Pines, Salinas Valley, the University of Pavia and the Kennedy Space Center.
For deployment, the model is quantised to 8-bit precision, fine-tuned using the Vitis-AI framework and integrated as a subordinate accelerator using AMD's Deep-Learning Processing Units (DPU) instance on the Zynq device.
Our tests show that the model can process each spectral band in an HSI image in 4\,ms, 2.6$\times$ better than INT8 inference on Nvidia's Jetson platform  \& 1.27$\times$ better than SOTA artefact detectors.
Our model also achieves an f1-score of 92.8\% and FPR of 0\% across the dataset, while consuming 21.52\,mJ per HSI image, 3.6$\times$ better than INT8 Jetson inference \& 7.5$\times$ better than SOTA artefact detectors, making it a viable architecture for deployment in CubeSats.

\end{abstract}
\input{Intro}
\input{background}
\input{architecture}
\input{experiments}

%
%
\bibliographystyle{ieeetr}
\bibliography{refs}

\end{document}

%% file: Intro.tex
\section{Introduction}

\begin{figure}{}
\centering
    \includegraphics[trim={2cm 0 1cm 0},width=0.8\linewidth,]{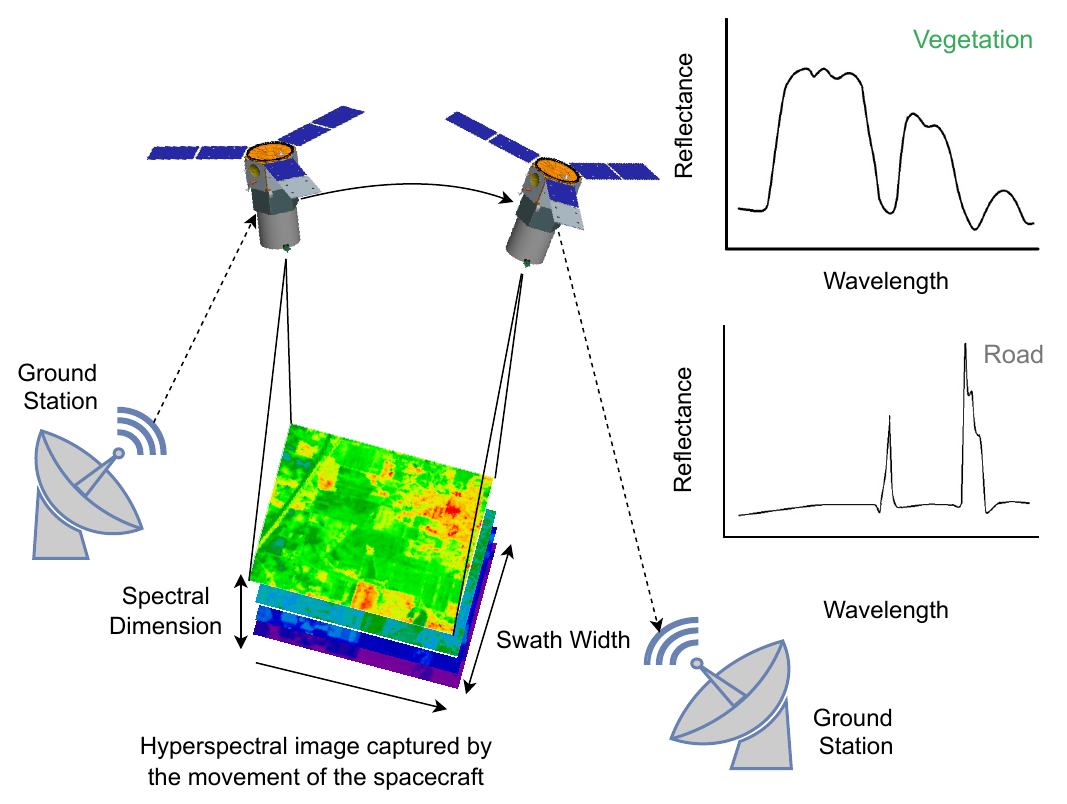}
    \caption{The concept behind imaging spectroscopy. An airborne or spaceborne satellite samples multiple spectral wavebands over a preset area. The hyperspectral image cube is then pre-processed with an onboard image processor before transmitting the cube image to a ground station.} 
    \label{fig:ImagingSpectroscopyConcept}
\end{figure}

Hyper-spectral imaging (HSI) provides the footwork to deeply investigate the Earth's atmosphere and its surface by capturing rich spectral and spatial information~\cite{grahn2007techniques}.
Compared to traditional multi-spectral imaging methods, HSI captures spatial images in continuous spectral regions, obtaining information on the spectral radiance of the area for different spectral bands.
Figure~\ref{fig:ImagingSpectroscopyConcept} illustrates the concept of how HSI images are captured for remote spaceborne satellites.

HSI imaging sensors are onboard most earth-orbiting satellites and provide valuable information for research and development in many fields such as agriculture, land cover, climate analysis, environment studies and industrial sectors \cite{agri2020}.
Images acquired by these sensors are typically transmitted to the ground station when the satellite enters its radio range.
These are subsequently processed offline and analysed to provide useful information to end users. 
Recently these systems have been increasingly incorporated into smaller satellites ($>$ 500Kg), especially \textit{cube satellites}~\cite{hypercube_v2,GHOSt,Hypso}. 
Cube-Satellites (Nano\{1Kg-10Kg\} and Micro\{10Kg-100Kg\}) offer multiple advantages in the form of reduced costs, higher risk distribution (operated in constellations) and higher data security among others, compared to conventional satellites which make them more accessible to a wider pool of companies~\cite{cubesatadvantages}. 

\begin{figure}{}
\centering
    \includegraphics[trim={4cm 3cm 3cm 3cm},width=0.8\linewidth,]{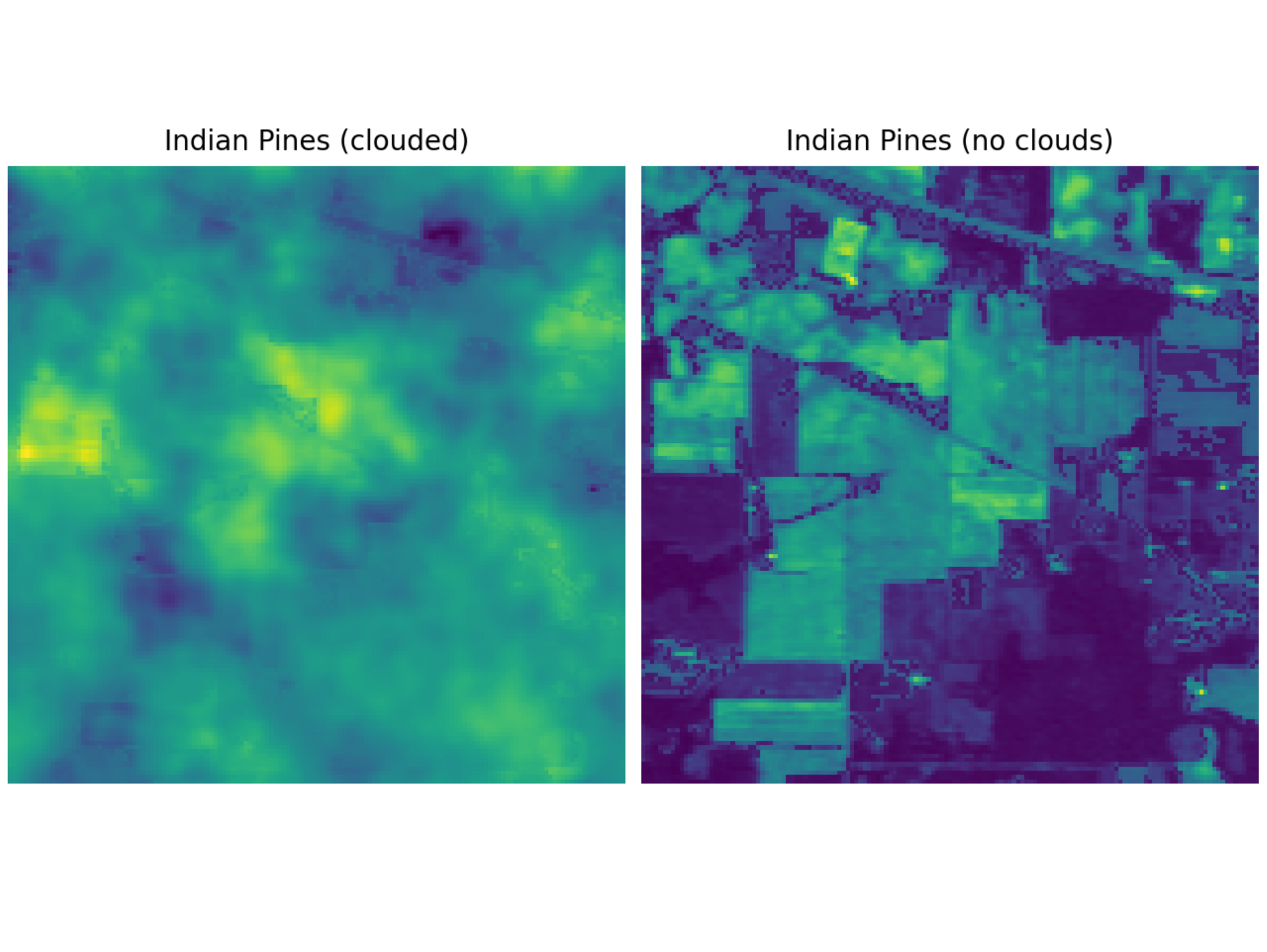}
    \caption{Example of a frame from an HSI imagecube with a cloud artefact}
    \label{fig:artefactExample} 
\end{figure}

Due to the orbiting nature of the satellites and the limited number of ground stations, there is a minimal window for these transferring HSI images to the ground station which takes up to 8\,GB on-board memory storage~\cite{Hypso}.
While downlink capabilities have increased over the past decade, the complexity and size of data acquired \& transmitted to the ground stations by modern HSI sensors have outpaced this increase. 
Another key concern with ground station-based processing is that HSI images acquired by mini- and CubeSats could contain artefacts such as clouds or distortions due to a malfunctioning sensor. 
Figure~\ref{fig:artefactExample} shows a cloud artefact within the HSI images that prevents extraction of useful vegetation information contained with the specific wavelength bands. 
Transmission of artefact-laden HSI leads to wasteful use of energy and communication bandwidth available with CubeSats with low energy budgets and tight communication windows.
Adding a neural network-based computation engine (pre-processor) onboard satellites to filter artefact-laden HSI images could prevent the wasteful use of precious communication resources.
Due to the general importance of this problem and its advantages, multiple \textit{artefact detectors} have been presented in the research literature based on convolutional neural networks (CNNs)~\cite{hu2022hyperspectral,lightweightAnomalydetection}.
Such CNN models have shown to be highly accurate in classifying artefacts in HSI images.
After deployment, these detectors offer an advantage in that they can be continuously improved with periodic weight updates to maintain operational accuracy, ensuring the transmission of only useful HSI images, and indirectly contributing towards extending the operating life of these satellites. 

Integration of ground-station-based HSI processing models for onboard satellites could be an option; however, even with their high accuracy, adapting them to satellites, especially CubeSats, with limited computing capabilities and energy budget is not considered feasible.
Most ground station-based HSI processing models proposed in the literature make use of large-scale deep learning models like dynamic routing~\cite{hybrid_CapNet} and capsule networks~\cite{deng2018hyperspectral} with complex layer structures (3D convolutional layer and group normalisation layer), rendering them impractical for satellite deployment.
Using computational models to improve energy- (by pre-processing HSI images on-board satellites) and communication efficiency of CubeSats are gaining popularity, as shown by authors in~\cite{phi-sat,imagedua2020comprehensive} respectively.
Model compression techniques such as quantisation are used to reduce the computational complexity (and thus energy consumption) of neural network models for deployment in space applications~\cite{ghiglione2022survey,wei2019fpga}.
Quantised deep-learning accelerators on Intel Movidius Vision Processing Units (VPUs) have been integrated into onboard HSI imaging systems for detecting artefacts within the captured images, before transmitting them to ground stations~\cite{cloudscout2020}.
Generalised FPGA-based acceleration of deep-learning models have also been shown to perform artefact detection using quantised neural networks (QNN), increasingly focussing on the energy-efficiency of these solutions~\cite{cloudsatnet,rapuano2021fpga}. 

For the specific problem of artefact detection in HSI images for on-board applications, most approaches rely on supervised learning-based DL models~\cite{cloudscout2020,rapuano2021fpga,cloudsatnet}. 
While they perform well in detecting specific artefacts (e.g. cloud cover), they do not generalise well to other capture-time artefacts (e.g. agricultural/flora spectral changes).
Developing artefact detection models that can generalise to detect the usefulness of captured HSI images, by detecting and classifying artefact-laden images, is hence of interest. 

In this paper, we propose an unsupervised 2D convolutional autoencoder network as an artefact detection model for on-board HSI processing. 
To enable deployment in a resource-constrained satellite environment, the model (weights, biases and activations) is quantised to 8-bit precision targeting a commercial off-the-shelf (COTS) XCZU7EV FPGA on the ZCU104 development platform from AMD/Xilinx.
The model is quantised using the Vitis-AI toolchain and deployed on an off-the-shelf deep learning processing unit accelerator~\cite{VitisAI,zynq}.
The network generalises to the complex features embedded in HSI images and learns to reconstruct them using unsupervised learning techniques.
It then concludes whether the image contains artefacts based on a custom reconstruction-loss metric. 
This gives the model the ability to generalise beyond regular cloud-clover and classify any capture-time anomaly in the HSI image as an artefact, making it more robust and increasing its overall efficiency.

The main contributions of this work are as follows:
\begin{itemize}
    \item We present a novel 8-bit quantised 2D convolutional autoencoder network for anomaly detection in HSI images with an f1 score of 92.8 \% and a 0 \% FPR achieving state-of-the-art accuracy when tested using multiple datasets.
    \item We present a custom reconstruction error function applied to the network's output to detect artefacts in the sensed image. 
    \item The quantised model is tested on the ZCU104 evaluation board and achieves a per-image processing latency of 4\,ms which is comparable to other works proposed in the research literature~(Section~\ref{results}).
    \item We present a comprehensive comparison against the SOTA and the NVIDIA Jetson Xavier GPU (INT8) and find the proposed model performs 1.27~$\times$, 2.6~$\times$ and 7.5~$\times$, 3.6~$\times$ better in terms of per image processing latency and per-inference energy consumption respectively.  
\end{itemize}

%% file: background.tex
\section{Background}

\subsection{HSI Classification \& Artefact Detection}

Hyperspectral images can be represented as a datacube of (x,y,w) dimensions, where \emph{x} and \emph{y} are the spatial dimensions of the image and \emph{w} represents the spectral dimension for spatial images taken in different wavelength ranges.
The problem of HSI classification lies in classifying individual pixels in a spatial region based on the available spectral information. 
Early models for HSI classification include SVM implementation~\cite{melgani2004classification} and unsupervised approaches like mean shift filtering~\cite{lee2010unsupervised}. 
CNN-based classification architectures such as the 2D autoencoder model described in~\cite{tao2015unsupervised2d} have been explored for feature extraction and dimensionality reduction with great success.
Napela et al. \cite{unsupervised3d} proposed a 3D unsupervised autoencoder network for segmenting images without known labels by learning the latent representation through the autoencoder and then applying clustering over the output. 
Khodadadzadeh et al.~\cite{hybrid_CapNet} proposed a deep complex neural network using an autoencoder with a capsule network to represent the classes of the pixels better. 
The current state of artefact/anomaly detection-based models reviewed comprehensively by Hu et al. \cite{hu2022hyperspectral} use SOTA models like Convolutions AutoEncoders (CAEs), Generative Adversarial Network (GAN), Recurrent Neural Networks (RNN), among others, that aim to classify which pixels contain anomalies. 
However, their approaches focused on searching for small occlusions such as planes, boats, buildings, and vehicles, and does not perform generalised anomaly detection for a large spatial region (e.g., a change in a large spectral region of a spatial location). 
Unsupervised models for HSI image classification on the other hand can rely on the large volume of HSI datasets for land cover found in USGS earth explorer~\cite{Additionaldataset} to train models. 
Post-training with clustering can be then implemented to classify the results as seen with the 3D unsupervised autoencoder model proposed in~\cite{unsupervised3d}.
While all discussed models build on the principle of pixel-accurate classification of classes or small anomalies, the high computational complexity of these proposed models limits their adoption for generalised anomaly/defect detection onboard satellites due to strict restrictions on power usage, storage and processing power.

\subsection{Related Works}

Onboard cloud detection plays a crucial role in transmitting data for land cover applications, where cloud cover in captured images is considered as an anomaly, and hence the image needs to be excluded.
Giuffrida et al.~\cite{cloudscout2020} introduced the CloudScout CNN model for classifying HSI cloudy images, achieving 92.3\% accuracy with a false positive rate of 1\% using a 70\% cloudiness threshold. 
A variation of this model, quantised at 16b-floating precision implemented on the Intel Myriad-2 VPU was subsequently deployed on the Hypserscout-2 Cubesat which completed multiple successful missions~\cite{HyperScout_v2}. 
Rapuano et al.~\cite{rapuano2021fpga} expanded upon the CloudScout model by implementing INT8 quantisation and deploying it on the Zynq UltraScale+ ZCU106 at a 0.3\% reduced accuracy trade-off while accelerating the inference by a factor of 2.4x. 
More recently, Pitonak et al.~\cite{cloudsatnet} enhanced cloud coverage capabilities by employing a deeper quantised CNN model with 10 layers, quantising the model to INT4 precision using the FINN toolchain, and testing it on the Z-turn board equipped with the Xilinx Zynq 7020 SoC.
Their improvements yielded higher accuracy in classifying cloudy images, along with a 2.2x increase in inference speed.
These studies show the effectiveness of quantised CNN models on FPGAs for satellite deployment.
These research outputs also provide insights into latency requirements and energy limits available for satellite systems in addition to the baseline accuracy metrics. 
While our study is similar in scope to the exploration done by Ma et al.~\cite{lightweightAnomalydetection}, in that both schemes use a stacked autoencoder (SAE) as an anomaly detector with 8-bit model quantisation and an FPGA-based SoC as the platform, our study diverges by focusing on developing a general artefact/anomaly detector where not only smaller anomalies but large anomalies such as unexpected spectral changes within agricultural field or flora changes within a spatial location can be successfully flagged. 

\begin{table}[htbp]
\centering
\caption{Summary of related works}
\scalebox{1.0}{
    \begin{tabular}{@{}lcl@{}}
    \toprule
    \textbf{Application} & \textbf{Spectral range} & \textbf{Model type} 
    \\
    \midrule
    \textbf{HSI classification} &  & \\
    Mean Shift Filtering (2010) \cite{melgani2004classification} & 400-2500 nm & PCA + MSF
    \\
    SSAE (2015) \cite{tao2015unsupervised2d}  & 430-860 nm & SSAE + SVM 
    \\
    3D CAE (2020) \cite{unsupervised3d} & 400-2500 nm & 3D CAE
    \\
    HCapsNet (2021) \cite{hybrid_CapNet} & 400-2500 nm & PCA +CAE + CapNet
    \\
    
    \midrule
    \textbf{Cloud Detection (Anomalies)} & & 
    \\
    CloudScout (2020) \cite{cloudscout2020} & RGB  & QCNN 
    \\
     CloudScout Extended (2021) \cite{rapuano2021fpga}& RGB & QCNN 
    \\
     CloudSatNet (2022) \cite{cloudsatnet} & RGB & QCNN 
    \\
    \bottomrule
    \end{tabular}
}
\label{table:dpuResourcecomparison}
\end{table}

%% file: architecture.tex
\section{System Architecture}
\label{system}
\subsection{Convolutional Auto-Encoder as an anomaly detector}

The design our generalised anomaly detector, our choice of convolutional autoencoders (CAE) as the neural architecture was inspired by the work in ~\cite{hybrid_CapNet}. The CAE is divided into 2 layers corresponding to an encoder $E_W(\cdot)$ and a decoder $D_U(\cdot)$. It aims to reconstruct the input data $x$ by extracting the latent features $h$ (eqn. \ref{eqn:encoder}) at the bottleneck and then reconstruct the image using the information present at the bottleneck (eqn. \ref{eqn:decoder}), while minimising the mean square error(MSE) between its input and output, i.e.

\begin{equation}
\label{eqn:AE}
\centering
Loss =  \min_{W,U} \frac{1}{n \cdot m}  \sum_{j=1}^{m} \sum_{i=1}^{n} | x_{i,j} - D_U(E_W(x_{i,j}))|^2
\end{equation}

For a fully connected convolutional autoencoder,

\begin{equation}
\label{eqn:encoder}
\centering
E_W(x) = \sigma(x \ast W) \equiv h 
\end{equation}

\begin{equation}
\label{eqn:decoder}
\centering
D_U(h) = \sigma(h \ast U)
\end{equation}

Where $x$ is the input tensor, $h$ is the latent tensor, $W$ and $U$ are the weights of the encoder and decoder respectively, "$\ast$" is the convolutional operator, $n$ and $m$ are the dimension of the input and $\sigma$ is the activation function like ReLU or sigmoid. 
Using the custom reconstruction error discussed later in section \ref{reconstruction}, we determine a threshold value to classify anomalies in acquired HSI images.

\subsection{Design Space Exploration for CAE parameters}
\label{design}
We explored multiple configurations of the CAE through a manual design space exploration arriving at the best-performing configuration using detection accuracy and FPGA resource estimates as the guiding factors. 
The exploration was also guided by the layers supported by the compilation tools from AMD for mapping the CAE architecture to a synthesisable/executable design. 

In the manual design search, we experimented with [4,6,8] layer-deep models. 
We observed the 6-layer-deep model best learnt the latent representation of the HSI imagecube based on the validation scores during training.
For feature map sizes in the \textit{Conv2D} and \textit{UpSample2D}+\textit{Conv2D} layers, we found that the [128,256,512] configuration for the encoder and [512,256,128] for the decoder gave the highest accuracy among the other configurations.
The encoder's input \textit{Conv2D} layer takes input from an HSI image of size 144x144. 
The subsequent \textit{Conv2D} layer processes the image with a stride of 2 reducing its dimensions to size 72x72. 
The next \textit{Conv2D} layer then processes the image with the same size to learn/extract valuable features.

\begin{figure}[t!]
\centering 
    \includegraphics[trim={3cm 0 2cm 0},width=0.9\linewidth,]{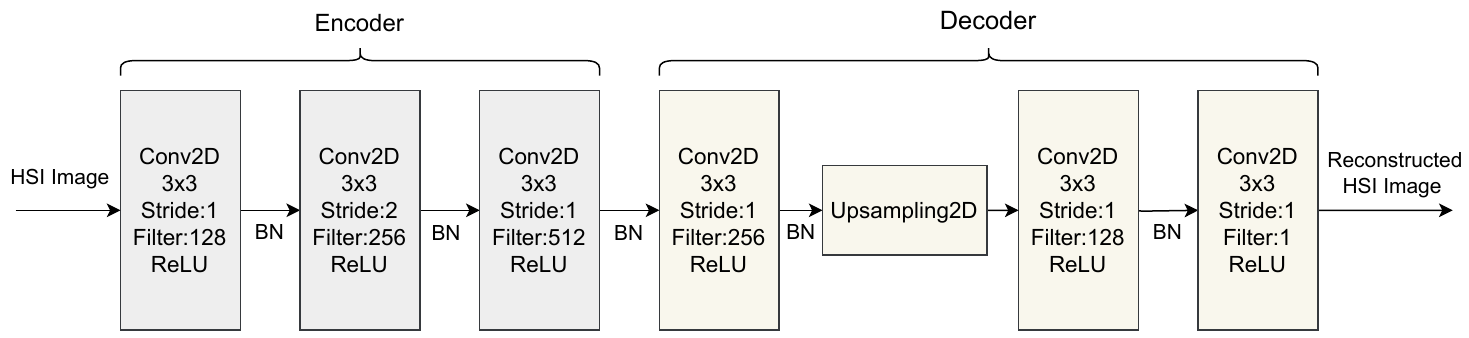}
    \caption{The proposed model convolutional autoencoder model as a defect detection.}
    \label{fig:ModelSummary} 
\end{figure}

\begin{figure}[t!]
\centerline{\includegraphics[width=70mm,scale=1.5]{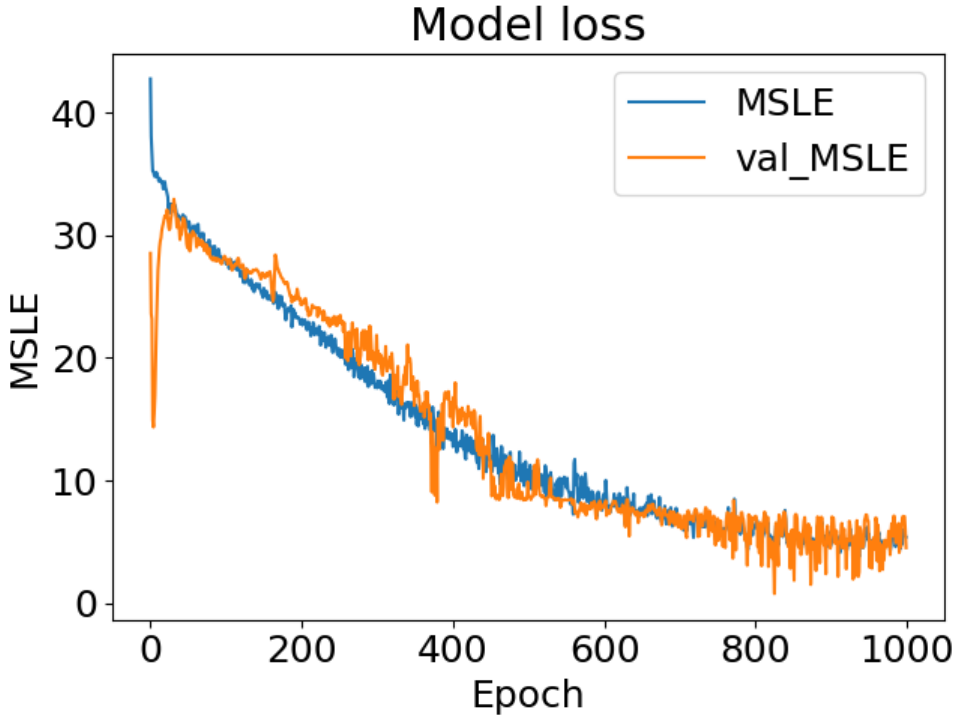}}
\caption{MLSE model loss per epoch.}
\label{loss} 
\end{figure}

The decoder is then used to increase the dimensionality of the bottleneck to reconstruct the input. 
The 72x72 image with 512 filters is then squished back into 256 filters with a \textit{Conv2D} layer. 
The image is restored to the original size using \textit{UpSampling2D} resulting in a 144x144 image with 128 filters.
This output is then linked to a \textit{Conv2D} layer with 1 filter acting as the output of the model as shown in figure~\ref{fig:ModelSummary}. 
The model is trained for 2000 epochs with no notable over-fitting or under-fitting as seen in figure~\ref{loss}.

\subsection{Dataset and Training}
\label{section:dataset}
While HSI datasets are widely available through international agencies such as the USGS, almost all 
sources provide unlabelled raw HSI frames. 
Most practical HSI analysis models previously discussed require the labelling of these HSI images for training the model. 
In this paper, we use widely known HSI scenes from Indian Pines (IP), Salinas Valley (SV) and Kennedy Space Center (KSC) that were captured using the Airborne Visible Infra-Red Imaging Spectrometer (AVIRIS) acquired by NASA. 
Additionally, the University of Pavia (UP) captured by the Rosis sensor is also used. These datasets will serve as a benchmark for testing of the model architecture and deployment into the XCZU7EV FPGA.
The spectral bands containing water absorption regions that predominantly capture clouds are removed and datasets like SV, UP \& KSC are divided in patches of 144x144 as inputs to the model. The final dataset specification is described in table~\ref{tab:my_label}. 
IP and SV patches are used to train the model in a 67/22/11\% train/test/validation split respectively. During validation, additional test data from UP and KSC is added to the 11\% validation data from IP + SV. The validation data is classified as "seen before" (1) and the additional UP + KSC data is classified as "not seen before" (0).

The proposed model architecture from section~\ref{design} comprises 2.95 million parameters with an input dimension of 144 x 144. 
The HSI cubes with larger dimensions like the Salinas Valley, University of Pavia and Kennedy Space Center are portioned into 144x144 regions for input into the model.
Adam optimizer with a learning rate of 0.001 was chosen for training. 
\textit{Mean Squared Error} (MSE) was used as the loss function.
To improve the convergence and prevent overfitting during training, \textit{BatchNormalization} layers were used between each Conv2D layer.
After training, the threshold of the model is determined using the custom reconstruction error from section~\ref{reconstruction} by testing the model against its held-out test dataset. 

\begin{table}[]
    \centering
    \caption{HSI Dataset specification}
    \begin{tabular}{@{}p{0.2\linewidth}rrr@{}}
         \toprule
         \textbf{Dataset} & \textbf{Spatial} & \textbf{Spectral} & \textbf{Spectral } \\
         & \textbf{size} & \textbf{bands} & \textbf{range (nm)} \\
         \midrule
         Indian Pines & 145x145 & 200 & 400-2500\\
         Salinas Valley & 512x217 & 204 & 400-2500\\
         University of Pavia & 610x610 & 103 & 430-860\\
         Kennedy Space Center & 512x614 & 224 & 400-2500\\
         \bottomrule
    \end{tabular}
    \label{tab:my_label}
\end{table}

\subsection{Custom Reconstruction Error after Inference}
\label{reconstruction}
We construct a combination of mean squared error MSE (\ref{MSE}) and mean squared logarithimic error MSLE (\ref{MSLE}) loss functions as the custom-loss function used to obtain the \textit{Reconstruction Error} ($R_{err}$) between the input and output HSI images.
The combined metric is the addition of both metrics with a fine-tuned factor multiplier \textit{k} for MSLE which is a hyperparameter for tuning. 
Evaluation of the error threshold is a vital step in splitting artefact-free \& artefact-laden HSI images. 
The reconstruction error that results in the highest f1 score of the model is chosen as the threshold boundary. 
The reference equations are shown below \ref{RE}.

\begin{equation}
\label{MSE}
MSE =  \sum_{i=1}^{144}\sum_{j=1}^{144} \frac{(x[i,j]-y[i,j])^2}{ n*m}
\end{equation}

\begin{equation}
\label{MSLE}
MSLE =  \sum_{i=1}^{144}\sum_{j=1}^{144} \frac{(log(x[i,j])-log(y[i,j]))^2}{ n*m}
\end{equation}

\begin{equation}
\label{RE}
R_{err} = MSE + k * MSLE 
\end{equation}

The ranges of intensities of the pixels contained within the HSI datacube described in \ref{section:dataset} vary by orders of magnitude (0-10000) over different bands of wavelengths. 
Hence, MSE alone can misguide results in case of outliers in the dataset.
Therefore, 
we also use MSLE loss function in conjunction with MSE as it treats outliers on the same scale as normal data due to logarithm's properties and also helps amplify smaller errors that MSE does not take into account. 
There are three cases of reconstruction loss possible: images where \textit{MSE} is dominant, images where both \textit{MSE} and \textit{MSLE} are equal and lastly, where \textit{MSLE} is dominant. For deployment, the aim is to find the value of \textit{k} where the \textit{average} contributions of both loss functions is similar. In our case, \textit{threshold} is obtained by finding a reconstruction error value that maximises the f1-score of the model.
Through extensive design-space exploration, a threshold of $\approx\,70000$ was found to maximise the f1-score across our training and validation sets.

\subsection{Hardware Generation and Integration with the CAE Model}

Once the model architecture was determined and trained, we utilised AMD's Vitis-AI framework to generate the accelerator with the hybrid Zynq Ultrascale+ as the target platform.
The Zynq Ultrascale+ device integrates a dual-core ARM real-time core and a quad-core ARM processor within the processing system (PS) section of the device. 
The programmable logic (PL) section of the device enables the addition of specialised custom logic blocks and accelerators which are accessible using the Advanced eXtensible Interface (AXI) protocol from PS. 
Vitis-AI generates an executable model file, which can be deployed using AMD's DPU IP block in the PL.
The DPU is an instruction-based array of programmable processing engines (PEs) for accelerating deep-learning inference on FPGAs.
The Vitis AI design flow compiles the Tensorflow model to executable instructions for the DPU's PE.

\begin{figure}
\centering
    \includegraphics[width=0.9\linewidth,]{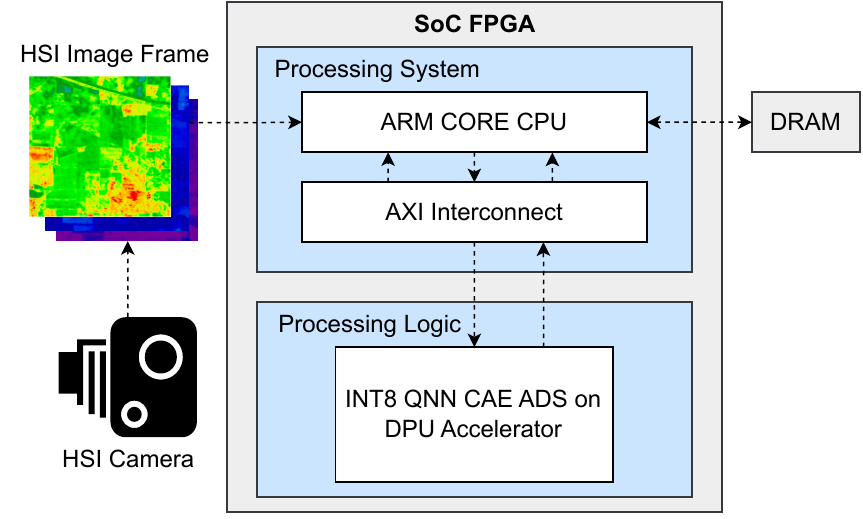}
    \caption{Proposed system architecture of the integrated ADS.
The quantised CAE model is accelerated on the PL part of the FPGA device.}
    \label{fig:systemArchSmall}
\end{figure}

Figure \ref{fig:systemArchSmall} shows the high-level system architecture for our proposed HSI imaging system on a Zynq platform, with our HSI analyser in the processing pipeline.  
The HSI camera is interfaced through the CPU which receives the image for processing, which is fed to the inference accelerator (our HSI analyser) via direct memory access (DMA).
As mentioned before, AMD's DPU IP runs the quantised CAE model to detect anomalies in the image. 
Standard Vitis AI Runtime (VART) APIs are used to configure and communicate with the DPU-based Artefact Detection System (ADS) engine.
The model uses interrupts indicating the completion of tasks on the PS, allowing for software tasks to run in a non-blocking fashion. 
This allows for a continuous flow of frames to process in the DPU.
For our evaluation, we used Linux OS with Petalinux tools to enable a seamless interface to the accelerator at the cost of slightly higher software complexity (compared to a bare-metal or real-time OS).

%% file: experiments.tex
\begin{table}[htbp]
\centering
\caption{Inference accuracy percentage metrics of the 2D CAE in different phases from FP32 to INT8 quantisation} 
\scalebox{0.9}{
\begin{tabular}{@{}lcccccc@{}}
\toprule
\textbf{Model} &  \textbf{Accuracy} & \textbf{F1} & \textbf{Recall} & \textbf{Precision} & \textbf{FNR} & \textbf{FPR}\\
\midrule

 Full precision CAE & 98.92 & 94.05 & 91.0  & 100.0   & 11.24  & 0.0\\
 8-bit QCAE & 90.4 & 0.0  & 0.0  & 0.0 &  100 & 0.0\\
8-bit finetuned QCAE & 98.70 & 92.77 & 88.76  & 100.0   & 13.48  & 0.0 \\

\bottomrule
\end{tabular}}
\label{table:accuracy} 
\end{table}

\begin{figure*}[htbp]
\centerline{\includegraphics[width=0.8\linewidth]{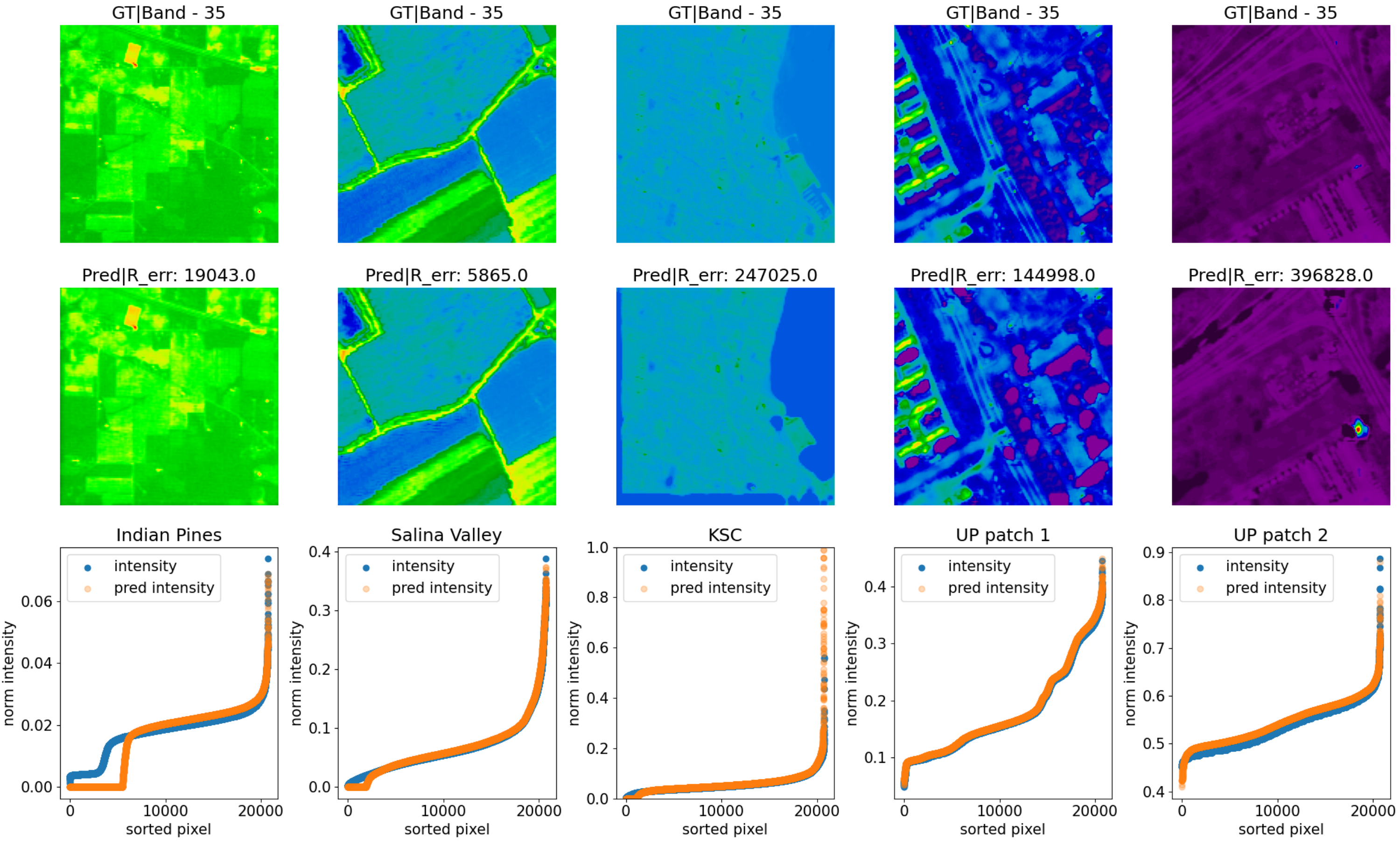}}
\caption{Shows image ground truth and prediction of frames from IP, SV, KSC and UP datasets. The last row shows the comparison of the spread of sorted pixels based on intensity between ground truth and inferred images.}
\label{fig:IP_pred} 
\end{figure*}

\section{Deployment and Experimental Results}
\label{results}

The CAE model is trained using an Nvidia Gforce RTX 3080-Ti GPU using the training, test and validation split discussed in section~\ref{section:dataset}.
The best model obtained from this training flow is exported to the Vitis-AI toolchain for the quantisation of model parameters, with a uniform 8-bit quantisation applied to all weights, biases and activations. 
Vitis-AI's post-quantisation fine-tuning flow was used with the same set of training/validation data to mitigate the slight accuracy loss incurred by the post-training quantisation. 
The fine-tuning was executed for 50 epochs and the model achieved a 92.77\% F1 score as seen in table \ref{table:accuracy}

The fine-tuned quantised model is then packaged as an `xmodel' file ready to be deployed on the ZCU104 SoC. 
This model is executed using the B4096 version of the DPU accelerator in the PL part of the XCZU7EV FPGA device (4096 number indicating highest parallelism attributes\{operations/per\_cycle\}).
The DPU is synthesized at a 300\,MHz interface clock and a 600\,MHz DSP core clock frequency.
Vitis-AI runtime (VART) libraries are used to facilitate data movement to \& fro from the DPU accelerator. 



Using a pre-baked reconfigurable processing engine like the DPU allows for constant updates to the model (with new model weights) maintaining or in some cases improving the artefact-detection process allowing it to execute robustly. 
 This allows us to not re-write the bitstream on the FPGA device which is $\sim$10$\times$ more in size ($\approx$ 23\,Mb for our specific Zynq device) and more computationally expensive than updating weights of the network proposed in the paper ($\approx$ 2Mb). This directly leads to savings in operating costs, while still allowing post-deployment updates to extend the service lifetime of the anomaly detection system.

To quantify the system's performance, we measure the accuracy, per-image inference latency and per-inference energy consumption and compare them against the state-of-the-art artefact detectors presented in the literature~\cite{cloudscout2020,cloudsatnet,rapuano2021fpga}.
We also compare the quantised model with the FP16 \& INT8 implementations of the CAE model on a Jetson Xavier NX module.
The Xavier NX architecture features 384 CUDA cores and 48 Tensor cores \cite{NvidiaXavierGPU} to accelerate deep learning tasks.
The GPU is operated in the 10\,W desktop mode for our tests, with 4 out of 6 CPU cores operating clocked at 1.9\,GHz while the Xavier NX GPU achieved a peak clock rate of 510\,MHz during inference. 

\begin{table}[b!]
\centering

\caption{Confusion matrix of different quantisation and deployment. }
    \scalebox{1}{
        \begin{tabular}{@{}p{0.3\linewidth}lrr@{}}
        \toprule
            \textbf{Model} & \textbf{Message Type}  & \multicolumn{2}{@{}c@{}}{Predicted} \\
            \cmidrule{3-4}
            & & Similar &  Anomalous \\
            \midrule
          \multirow{2}{*}{FP32 CAE} & True Similar  &  79  & 10     \\
            & True Anomalous    & 0   & 837    \\
            
           \midrule
            \multirow{2}{*}{Vitis AI INT8 QCAE} & True Similar    &    77   &   12   \\
           &  True Anomalous    &    0   &  837    \\
           \midrule
            \multirow{2}{*}{Pytorch INT8 QCAE} & True Similar    &    81   &   8   \\
           &  True Anomalous    &    0   &  837    \\
            \bottomrule
        \end{tabular}}
\label{table:confmatrix}
\end{table}

\begin{table}[b!]
\centering
\caption{Comparison of inference accuracy against related works in the literature.} 
\scalebox{1}{
\begin{tabular}{@{}lcccccc@{}}
\toprule
\textbf{Model}& \textbf{Precision} &\textbf{Platform}&  \textbf{Accuracy} & \textbf{FPR}\\
\midrule

 CloudScout\cite{cloudscout2020}&16b-float&Myriad-2 VPU &92.3 &1.03\\
CloudScout ext.\cite{rapuano2021fpga}&8b-int& ZCU106 &92.0&-\\
CloudSatNet\cite{cloudsatnet}&4b-int&Zynq-7020&94.84&2.23\\
Proposed CAE&8b-int&ZCU104& 98.70 & 0.00 \\
\bottomrule
\end{tabular}}
\label{table:Stateaccuracy}
\end{table}

\subsection{Model Performance}
The detailed accuracy metrics of the full-precision model and the quantised version are presented in table~\ref{table:accuracy}. 
The fine-tuned quantised model achieves an F1 score of 92.77 \% slightly lower than the full-precision model which achieved an F1 score of 94.05\%, a loss of  $<$ 1.5\% due to the model quantisation. 
We also benchmarked the performance of the model on an Nvidia Jetson Xavier platform that also supports INT8 quantisation. 
For the same training time of 2000 epochs, the INT8 PyTorch model used for testing on the Xavier NX achieved a slightly better F1 score at 95.29\%. 
We can attribute this difference in performance to the different training and quantisation frameworks used by AMD (Vitis-AI) and Nvidia. 
We also present the confusion matrix of the models in table \ref{table:confmatrix} and see the model's classification strength in the identification of anomalous HSI images that contained artefacts. 


A sample of comparison between ground truth and the predicted image by the model is shown in figure \ref{fig:IP_pred}. 
We observe that for the IP and SV images (first two images from the left) the reconstruction error is below the set threshold of $\approx$70000 previously mentioned in \ref{section:dataset}. Whereas, the remaining KSC and UP patches (third, fourth and fifth images from the left) are above the set threshold providing insight that the model is able to differentiate between new images and images used during training.
The predicted IP \& SV images are almost identical. However, for the KSC and UP images, slight distortions occur due to the inability of the model to reconstruct unseen input HSI images faithfully.
The thresholds were chosen to achieve zero false positive detections (0\% FPR) since we want the model to always detect artefacts in acquired images; however, this leads to a higher false negative rate (13.48\% FNR in our case). 
We believe that this is an acceptable trade-off as the model would only transmit 13.48\% of overall artefact-laden images that it observes leading to the transmission of a high percentage of useful data from a CubeSat to the ground stations for further processing. 

We compare the inference performance of our CAE model against the state-of-the-art artefact detectors proposed in the research literature: CloudScout (Float16 model on Intel VPU )~\cite{cloudscout2020}, CloudScout Extended( INT4 model generated with FINN deployed on AMD/Xilinx FPGA )~\cite{rapuano2021fpga} \& CloudSatNet (custom quantised model implementation on Xilinx FPGA )~\cite{cloudsatnet} as shown in  Table~\ref{table:Stateaccuracy}, comparing them in terms of accuracy and FPR.
In terms of accuracy, our proposed model performs better than all the SOTA artefact detectors by 6.4\%, 6.7\% \& 3.8\% respectively.
In terms of FPR, the proposed model is better than~\cite{cloudscout2020} \&~\cite{cloudsatnet} by 1.03\% \& 2.23\% respectively, while FPR is not reported for the CloudScout Extended model.

\begin{table}[t!]
\centering
\caption{Performance of the QCAE on the B4096 DPU configuration along with other models.}
\scalebox{0.9}{
\begin{tabular}{@{}l@{ }cccc@{}}
\toprule
\multicolumn{5}{@{}c@{}}{\textbf{Impact on per frame latency (ms), FPS \& power consumption (W)}}\\\midrule
\textbf{Accelerator} & \textbf{Latency} & \textbf{FPS} & P\textsubscript{idle} &  P\textsubscript{tot}  \\ \midrule
Ours (ZCU104,INT8)          & 4  & 250                & 4.6   & 5.38  \\
Ours (Xavier NX,INT8) & 10.36 &96.52&2.2&7.6\\
Ours (Xavier NX,FP16) & 19.09 &52.38&2.2&8.2\\
CloudScout~\cite{cloudscout2020}  & 325 & 3.07&-&1.8 \\
CloudScout Extended~\cite{rapuano2021fpga}  & 141.68 & 7.05&-&3.4 \\
CloudSatNet~\cite{cloudsatnet} & 64.68 & 15.46&2.3&2.5 \\
\toprule
\multicolumn{5}{@{}c@{}}{ \textbf{\% Resource Utilisation on the platforms}}\\\midrule
\textbf{Accelerator} & \textbf{LUT} & \textbf{FF} & \textbf{DSP} & \textbf{BRAM / URAM} \\\midrule
Ours (XCZU7EV, INT8)            &  27.17&25.03&40.74&35.42 / 47.92  \\
CloudScout Ext~\cite{rapuano2021fpga} (XCZU7EV) &23.09& 3.79&67.30 &21.79 / 56.25 \\
CloudSatNet~\cite{cloudsatnet} (Zynq-7020) &69.05& 48.61&13.64 & 62.5 / -\\
\bottomrule
\end{tabular}}
\label{table:dpucomparison}
\end{table}

\begin{figure}[t!]
\centering\includegraphics[width=0.8\columnwidth]{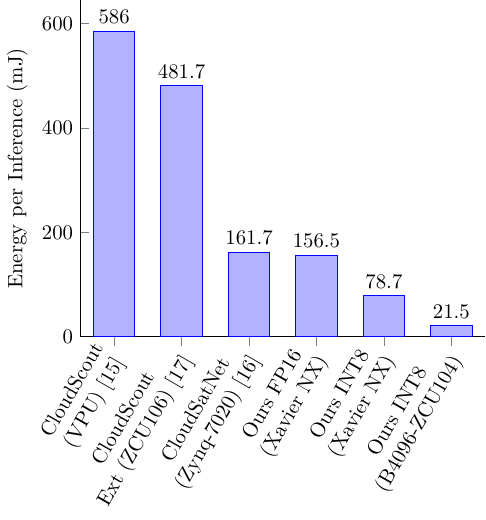}
\caption{Energy consumption per inference (E\textsubscript{inf}) comparison of our model v/s the state-of-the-art.
}
\label{fig:energy} \vspace{-10pt}
\end{figure}

\subsection{Latency, Power and Resource Consumption}
We quantified the latency of the quantised CAE model for each inference operation and observed this to be an average of 4.0\,ms averaged across 1000 runs.
This leads to a throughput of 250 HSI images per second, which is much higher than the typical acquisition rate of satellite imagery systems. 
Table~\ref{table:dpucomparison} compares our results against other approaches in the literature, that utilise different platforms (Xilinx's ZCU106 \& Zynq-7020 FPGA devices). 
Compared to the state-of-the-art CloudSatNet model~\cite{cloudsatnet}, our model achieves a 1.27$\times$ improvement in per-inference processing latency at identical image frame patch size.
The higher throughput also allows our model to dynamically scale inference performance, trading off inference throughput for energy consumption through simple frequency scaling schemes. 
We also quantified the per-inference latency of the model on a Jetson Xavier NX GPU using the \textit{torch2trt} framework that acts as a wrapper and utilises Nvidia's \textit{TensorRT} framework to perform inference with reduced precision (FP16 \& INT8) on GPUs. 
We observe that our model implementation is 4.7$\times$ \& 2.6$\times$ faster than the FP16 \& INT8 implementations on the Jetson Xavier NX GPU respectively, as shown in Table~\ref{table:dpucomparison}. 
We quantify the energy consumption of the FPGA monitoring PYNQ-PMBus power rails during inference. 
The active power consumption was observed to be 5.38\,W, which leads to a per-inference energy consumption of 21.52\,mJ.
Compared to state-of-the-art detectors, our model shows a 7.5$\times$ improvement in per-inference energy consumption~\cite{cloudsatnet}.
We also quantify the energy consumption of the FP16 \& INT8 variants of the Jetson device using the \textit{jtop} utility. 
We find that the proposed model is 3.6$\times$ \& 7.2$\times$ more energy efficient than the INT8 \& FP16 implementations of the model on the Jetson device.
Figure~\ref{fig:energy} shows the energy-per-inference comparison of the proposed model with the state-of-the-art detectors and the Jetson device implementations along with their classification accuracies.

In terms of resource utilisation, our model consumes slightly higher general purpose resources (LUTs, FFs, BRAMs) in place of lower DSP and URAM compared to the CloudScout Extended model which uses the same FPGA device (XCZU7EV-2FFVC1156) on the ZCU106 platform. 

\section{Conclusion}
In this paper, we proposed an unsupervised learning-based 2D convolutional autoencoder model integrated as an FPGA-accelerated energy-efficient model for onboard HSI image analysis in a satellite imagery system. 
We used openly available HSI datasets to train and fine-tune the model and deployed them using AMD's Vitis-AI framework using a DPU IP block on the Zynq Ultrascale+ platform. 
Our evaluation shows that the model can achieve an F1-score of 92.77\%  with an FPR of 0.0\% when trained on two distinct HSI datasets (IP and SV) and effectively determining artefacts in withheld datasets (UP and KSC) datasets.
We also achieve 1.27$\times$ improvement in latency and 7.51$\times$ improvement in energy per inference compared to state-of-the-art models in the literature, making our solution more appealing for satellite onboard processing. 

The solutions we have presented have some limitations due to unbalanced test dataset distributions, which could be improved in future works by equalizing the distributions and also exploring additional datasets from USGS~\cite{Additionaldataset}. 
In the future, we aim to integrate additional datasets to train and test our model against, as well as explore lower precision arithmetic for the model to improve the scalability of our approach.

